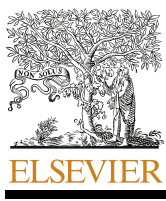
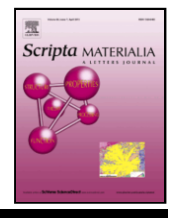

# Parameter-free prediction of phase transition in $PbTiO_3$ through combination of quantum mechanics and statistical mechanics

Zi-Kui Liu [a],[*], Shun-Li Shang [a],[*], Jinglian Du [a],[b], Yi Wang [a]

[a] *Department of Materials Science and Engineering, The Pennsylvania State University, University Park, Pennsylvania 16802, USA*
[b] *State Key Laboratory of Solidification Processing, Northwestern Polytechnical University, Xi'an, Shaanxi 710072, China*

ABSTRACT

Thermodynamics of ferroelectric materials and their ferroelectric to paraelectric (FE-PE) transitions is commonly described by the phenomenological Landau theory and more recently by effective Hamiltonian and various potentials, all with model parameters fitted to experimental or theoretical data. Here we show that the zentropy theory, which considers the total entropy of a system as a weighted sum of entropies of configurations that the system may experience and the statistical entropy among the configurations, can predict the FE-PE transition without fitting parameters. For $PbTiO_3$, the configurations are identified as the FE configurations with 90- or 180° domain walls in addition to the ground state FE configuration without domain wall. With the domain wall energies predicted from first-principles based on density functional theory in the literature as the only inputs, the FE-PE transition for $PbTiO_3$ is predicted showing remarkable agreement with experiments, unveiling the microscopic fundamentals of the transition.

© 20XX

It is challenging to computationally predict the experimentally measured properties of materials because only one or a few configurations are typically considered in computational approaches, whereas experimental measurements stem from sampling all possible configurations at all scales simultaneously. This challenge becomes acute for systems with phase transitions, which is often where the most fascinating properties exist such as ferroelectric materials discussed in the present work. Ferroelectric materials are with two or more discrete states of different nonzero electric polarization in zero applied electric field, referred to as "spontaneous" polarization, and can switch between these states with an applied electric field, stress, or temperature [1]. These effects are maximal near the boundary between polar ferroelectric (FE) and macroscopical non-polar paraelectric (PE) phases [2], commonly called a morphotropic phase boundary (MPB). They are traditionally described by the phenomenological Landau-Ginsburg-Devonshire theory (LGDT) [3] with the systematic modeling work reported by Cross and co-workers [4] by fitting model parameters to experimental observations.

Even though the LGDT formalism describes the FE and PE behaviors well macroscopically, it does not correctly describe the microscopic features, such as the domain walls (DWs) [5,6]. As a symmetry-based analysis of equilibrium behavior near a phase transition, LGDT postulates that the macroscopic polarization changes smoothly from a finite value in the FE phase to zero in the PE phase in accordance with macroscopic experimental measurements. On the other hand, the modern theory of polarization (MTP) developed more recently [1,7] focuses on differences in polarization between two different states, which is what actually measured in a polarization-reversal experiment. Therefore, the macroscopic polarization in the macroscopic FE phase (termed as system in the present work), defined as the sum of the dipole moments in a given cell divided by the cell volume in the LGDT formalism, does not reflect the realistic microscopic configurations (termed as configuration in the present work) with a sizeable fraction of the electronic charge being shared among ions in a delocalized manner. One critical result of MTP is that the microscopic polarization does not necessarily vanish for a centrosymmetric phase, i.e., the PE phase [1,7,8]. In efforts to predict the FE-PE transitions at finite temperatures, effective Hamiltonian and various potential approaches have been developed with model parameters fitted to first-principles calculations based on density functional theory followed by Monte Carlo or molecular dynamic simulations [9–13]. The predictions from those simulations presented a remarkable agreement with experimental observations in terms of the phase stability sequence, transition temperatures, latent heats, and spontaneous polarizations, providing insights into the order-disorder versus displacive character of the transitions and the importance of various interaction terms in the effective Hamiltonians [9,10,14,15]. However, the fitting of the energy surface in terms of a Taylor expansion and the to-be-predicted structure reduces the predictability of the existing approaches.

In the present work, our approach without any fitting parameters beyond DFT, developed for the prediction of magnetic transitions and recently termed as the zentropy theory [16–18], is utilized to predict the FE-PE transition in PbTiO$_3$ with all input data predicted from the DFT-based first-principles calculations in the literature.

The details of the zentropy theory can be found in our recent publications [16–18] and are briefly presented as follows. In statistical mechanics, the partition function of a system, $Z$, is the sum of the participation functions of all its configurations, $Z_k$. For a canonical ensemble under constant mass, volume, and temperature (NVT), $Z$ is related to the Helmholtz energy of the system, while $Z_k$ is typically related to $E_k$, the total energy of the configuration $k$ with the same NVT as the system. This means that each configuration is a pure quantum state with zero entropy [16–18]. In applying statistical mechanics to real systems, various types of coarse-graining approaches are implemented, resulting in configurations that are no longer pure quantum states. The total entropy of the system thus is the weighted sum of the entropies of individual configurations and the Gibbs entropy among configurations in terms of the Gibbs-Boltzmann distribution as follows

$$S = \sum_{i=1}^{N} p_k S_k - k_B \sum_{i=1}^{N} p_k \ln p_k \quad (1)$$

where $S_k$ and $p_k = Z_k/Z$ are the entropy and probability of configuration $k$, respectively, and $N$ is the total number of configurations including multiplicity of each configuration.

From Eq. 1, the partition function and Helmholtz energy of a system under constant NVT is written as follows [16],

$$Z = e^{-\frac{F}{k_B T}} = \sum_{k=1}^{N} Z_k = \sum_{k=1}^{N} e^{-\frac{F_k}{k_B T}} \quad (2)$$

$$F = \sum_k p_k F_k + k_B T \sum_k p_k \ln p_k \quad (3)$$

where $F$ and $F_k$ are the Helmholtz energies of the system and the configuration $k$, respectively. For pure quantum state configurations with $S_k = 0$, one has $F_k = E_k$, and Eq. 1 to Eq. 3 reduce to the commonly used statistical mechanics formalism [19].

The usefulness of the zentropy theory resides on that $F_k$ of the stable and metastable configurations of a system can be reliably predicted by first-principles calculations based on the density functional theory (DFT) [20,21], enabling the parameter-free prediction of Helmholtz energy of the system in contrast to the existing approaches that require additional models and fitting of model parameters [9–15]. The zentropy theory has been successfully applied to a number of systems with magnetic transitions [22–30], demonstrating remarkable agreement with available experimental data, particularly the positive and negative divergences of thermal expansion for Ce and Fe$_3$Pt at their respective critical points [31]. In those works, the configurations of magnetic materials are represented by ergodic co-linear magnetic spin arrangements [23], and the number of configurations equals to $2^n$ with $n$ being the number of atoms with spin in the supercell. In the present work, the zentropy theory is extended to the prediction of the FE-PE transition in PbTiO$_3$, and its configurations are discussed as follows.

PbTiO$_3$ is a simple ferroelectric material with a phase transition from the FE tetragonal structure to the PE cubic structure at about 763 K without external stress and electric field based on the X-Ray diffractions [32], and the cubic structure is unstable at 0 K based on the DFT-based first-principles calculations [33]. The volume of the tetragonal structure decreases with the increase of temperature, and the volume of the cubic structure increases with temperature [32]. However, when both Pb and Ti edges were measured via XAFS (X-ray-absorption fine structure) analysis with the time and spatial resolutions being ~10$^{-16}$ seconds (0.1fs) and 1st to 4th nearest neighbor shells, respectively, it was found that the displacements of both Pb and Ti atoms within the unit cell vary little with temperature below the transition and decrease only slightly above the transition temperature [15].

Fang et al. [34] performed *ab initio* molecular dynamics (AIMD) simulations for PbTiO$_3$ using the lattice parameters measured by X-ray in the literature [32]. In contrast to the conventional molecular dynamics (MD) analyses where results are averaged over time, Fang et al. [34] categorized the atomic configurations as a function of time in terms of the Ti-O bond lengths within the nearest-neighboring shell. It was observed that an appreciable amount of cubic configuration, i.e., equal length of all Ti-O bonds, exists at temperatures about 300 K, much lower than its FE-PE transition temperature of 763 K, even though the time-averaged overall atomic configuration is tetragonal. By following the time evolution of each Ti-O bond (see the video in Supplemental Material), it is observed that the cubic configuration originates from the switch of the c-axis of the tetragonal configuration from z direction to x or y direction through thermal fluctuations and vice versa. As the temperature increases, the switch becomes more frequent, resulting in an increase of thermal population of the cubic configuration and a decrease of macroscopic tetragonality.

It is striking to see that all local Ti-O environments are tetragonal except those that are switching from one tetragonal orientation to another tetragonal orientation, even above the FE-PE transition, with the supercell lattice parameters in the AIMD simulations fixed to the experimentally measured cubic lattice parameter [32], resembling the polar clusters discussed in the literature [35]. At the same time, these tetragonal orientation fluctuations generate local DWs. The simulated local lattice parameters and Ti displacements are in excellent agreement with those measured experimentally [15]. It is thus evident that the phenomenological LGDT formalism does not capture the fundamental physics behind the FE-PE transition in PbTiO$_3$, i.e., the switch between FE configurations that results in the macroscopic PE cubic structure characterized by the X-ray scattering with its time resolution much lower than those during the switch between the FE configurations.

In applying the zentropy theory [16] to predict the FE-PE transition of PbTiO$_3$, the configurations thus include the tetragonal FE with and without DWs. Based on the experimental and computational results in the literature, there are two types of DWs in PbTiO$_3$, i.e., 90° and 180° DWs as twins on the (101) and (100) planes, respectively [5], denoted by 90DW and 180DW in the present work, respectively, as shown in Fig. 1. The tetragonal FE without DW is the ground state, denoted by FEG in the present work. The polarizations on either side of the DW are almost perpendicular to one another across the 90° DW, and parallel with opposite orientation across the 180° DW due to the additional constraint that the normal component of the polarization to the DW should be continuous across the DW so that no net interface charge is present [5]. The highly charged head-to-head or tail-to-tail 180° DWs due to surface or other defects [36,37] are not considered in the present work due to their high DW energies [37], thus vanishing thermal probabilities without additional external constrains such as strains from a substrate.

Among three types of configurations for PbTiO$_3$, their multiplicities need to be considered. For the FEG configuration, the first domain has six energetically equivalent orientations with the c-axis direction along the $\pm x$, $\pm y$, or $\pm z$ axes, while the second domain can be placed on one of the two sides of the remaining two axes with its polarization in the same direction as the first domain, resulting in a total of $6 \times 2 \times 2 \times 1 = 24$ equivalent configurations. The same number of equivalent configurations exists for the 180° DWs except the polarization of the second domain in the opposite direction of the first domain, i.e., $6 \times 2 \times 2 \times 1 = 24$ configurations. For the 90° DWs, there are four additional degrees of freedom for the polarization direction in the second domain, i.e., two directions along each of the two remaining axes, resulting in $6 \times 2 \times 2 \times 4 = 96$ configurations. It is noted that the DW orientation in the 90° DWs changes with the polarization direction of the second domain in order to keep the DW free of net interface charge [36,37]. Considering that there is only one ground state configuration at 0 K fully defined by its electron density distribution in accordance with DFT [20,21], the multiplicity ratio of the FEG, 90DW and 180DW configurations is thus 1:4:1.

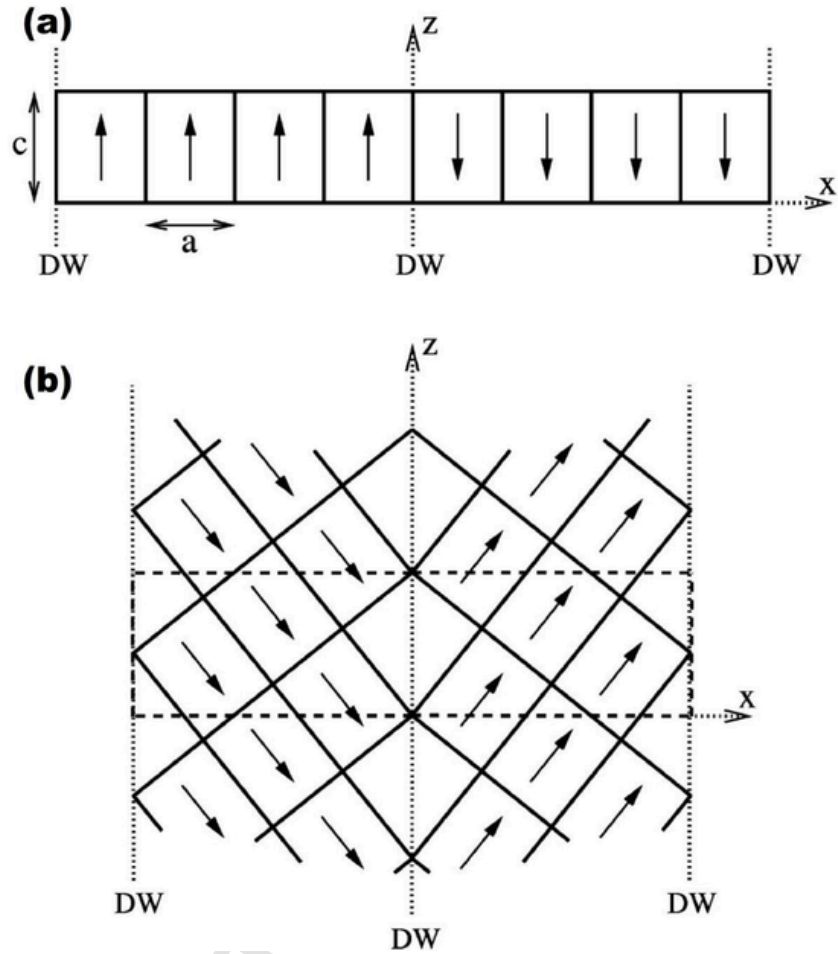

**Fig. 1.** Domain wall (DW) structures: (a) 180° DW and (b) 90° DW from [5].

The properties of ferroelectric DWs have been extensively studied in the literature [38]. The 90° DW energy in $PbTiO_3$ was predicted to be 35 and 29 mJ/m² using ultrasoft pseudopotentials (USPP) [5] and the projector augmented wave (PAW) method as implemented in VASP [39], respectively, both with the exchange-correlation functional being the local density approximation (LDA). It was pointed out that the PAW method slightly underestimates the spontaneous polarization compared with the USPP method [40], thus probably lower DW energies. The 180° DW can be Pb or Ti centered with the latter being unstable at a saddle point [5,41]. The Pb-centered 180° DW energy is predicted to be 150 or 270 mJ/m² using the norm-conserving non-local Troullier–Martins type pseudopotentials with or without inversion symmetry, respectively [41], 132 mJ/m² in terms of the high-symmetry centrosymmetric structure with inversion symmetry [5], and 128 or 114 mJ/m² with atomic relaxations in the z-axis direction only or in all the three spatial directions, respectively [42], using the LDA and PAW methods as implemented in VASP.

Unfortunately, no free energy of any DWs in $PbTiO_3$ has been reported in the literature. In principles, they can be predicted in terms of either phonon or Debye methods [43–45]. On the other hand, the present work uses all data from the literature in order to more objectively test the zentropy theory. As shown below, a remarkable agreement between the present predicted and experimental FE-PE transition temperatures is obtained though the agreement may be further improved with the entropies of all configurations included.

In the present work, two sets of 90° and 180° DW energies are used, i.e., (I) 35 and 132 mJ/m² from [5] with the interfacial areas in the supercell being $a^2\sqrt{1+(c/a)^2}$ and $ca$ with lattice parameters of $a = 3.86$ Å and $c = 4.04$ Å, respectively; and (II) 29 and 114 mJ/m² from [40] and [42] with $a = 3.8731$ Å and $c = 3.9990$ Å, respectively. The supercells used in all calculations contain 10 perovskite unit cells (total 50 atoms) and two DWs. The predicted probability of configurations using the two sets of DW energies is plotted in (need to combined with the next line into one paragraph)

Fig. 2 with the multiplicity taken into account, i.e., $4p_{90DW} = 4 Z_{90DW}/Z$ for the probability of the 90DW configuration (see the data file used for the calculations in Supplemental Material). Using 50% probability of the FEG configuration as the criterion based on our previous predictions for magnetic transitions [22–30] and the AIMD simulations [34], the predicted FE-PE transition temperatures are 776 K and 645 K for the dataset (I) and (II), respectively, which are 13 K above and 118 K below the experimental value of 763 K [32]. It is worth mentioning that the 50% probability is one of the simple criteria to determine phase transition, similar to the percolation threshold for a simply square system [46]. In addition to the 50% criterion, other criteria have been used to determine phase transition in our previous publications, including (i) the peak value of heat capacity (see the predicted Curie temperature in bcc Fe [25]) and (ii) the inflection point on the degree-of-disorder curve as a function of temperature (see the predicted Curie temperature in Invar alloy $Fe_3Pt$ [18]). All those criteria give similar phase transition temperatures.

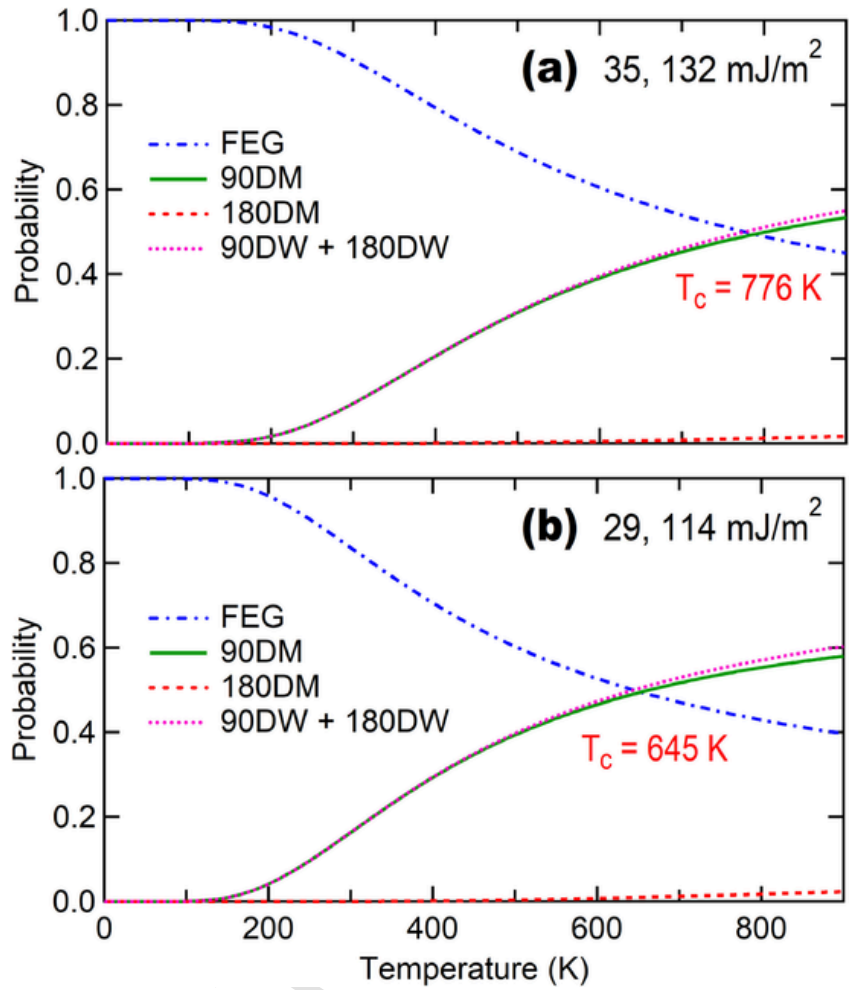

**Fig. 2.** Probability of configurations plotted as a function of temperature using datasets (I) and (II) with different combinations of DW energies: (a) 35 mJ/m$^2$ (90DW) and 132 mJ/m$^2$ (180DW) and (b) 29 mJ/m$^2$ (90DW) and 114 mJ/m$^2$ (180DW). The T$_c$ values are the predicted FE-PE transition temperatures.

Considering that only the DW energies at 0 K from DFT-based first-principles calculations in the literature are used, the present predictions without fitting parameters are remarkable. Our recent work in YNiO$_3$ showed that the magnetic transition temperature increases from 88 K to 144 K with the use of free energies for individual configurations in comparison with the use of total energy, which is with much better agreement with experimentally measured 145 K [30]. This indicates that the difference among entropies of configurations in PbTiO$_3$ may be pretty small in comparison with those in YNiO$_3$.

With the DW energies at 0 K from the literature used in the present work, only the second summation in Eq. 1 can be evaluated, resulting in the entropy per supercell among configurations as follows

$$S^{conf} = -k_B \left( p_{FEG} ln p_{FEG} + 4 p_{90DW} ln p_{90DW} + p_{180DW} ln p_{180DW} \right) \quad (4)$$

The configurational entropies for the two datasets thus obtained are plotted in Fig. 3, showing higher values with the dataset (II) than those with the dataset (I) due to the higher probabilities of 90DW and 180DW configurations with the dataset (II), thus lower transition temperature. The evaluation of the total entropy of the system requires the entropies of all individual configurations, which is under further investigations. The better agreement with experimental data using the dataset (I) is also due to the same NVT used for all configurations in the DFT-bases first-principles calculations, as required by the statisical mechanics.

In summary, the zentropy theory is applied to predict the FE-PE transition in PbTiO$_3$ with the 90° and 180° DW energies determined by DFT-based first-principles calculations in the literature without fitting parameters, demonstrating remarkable agreement with experimentally reported FE-PE transition temperature. The present prediction from the zentropy theory captures the fundamental physics behind the FE-PE transitions in terms of thermal fluctuation of polarization directions and thus provides a theoretical guidance for the discovery and design of FE materials with emergent functionalities. Future work on PbTiO$_3$ includes the predictions of free energies of the FEG, 90DW, and 180DW configurations, and other properties such as heat capacity and negative thermal expansion of PbTiO$_3$ using our mixed-space approach for phonon calculations of polar materials [47]. Qualitatively, the negative thermal expansion in PbTiO$_3$ can be explained from the increased probability of the 90DW configuration which has its volume smaller than that of the FEG configuration [48] based on the zentropy theory applied to Fe$_3$Pt [24,31].

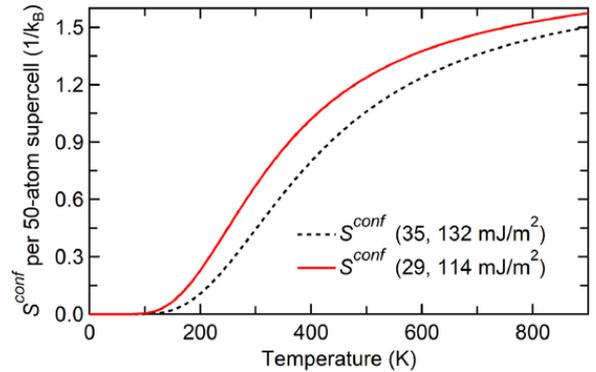

**Fig. 3.** Configurational entropy plotted as a function of temperature using datasets (I) and (II) with different combinations of DW energies.

**Declaration of Competing Interest**

The authors declare that they have no known competing financial interests or personal relationships that could have appeared to influence the work reported in this paper.

**Acknowledgments**

The authors acknowledge the support from the Endowed Dorothy Pate Enright Professorship at the Pennsylvania State University and the U.S. Department of Energy under Contract No. DE-SC0023185. The government reserves for itself and others acting on its behalf a royalty-free, non-exclusive, irrevocable, worldwide license for Governmental purposes to publish, distribute, translate, duplicate, exhibit and perform this copyrighted paper. JLD is financially supported by the China Scholarship Council during her visit at the Pennsylvania State University. The authors thank Huazhi Fang for providing the video from the AIMD simulations for PbTiO$_3$.

**Supplementary materials**

Supplementary material associated with this article can be found, in the online version, at doi:10.1016/j.scriptamat.2023.115480.